\documentclass{INTERSPEECH2023}
\usepackage{tabularx}
\usepackage{makecell}
\usepackage{array}

\newcolumntype{Y}{>{\centering\arraybackslash}m{0.232\textwidth}} 
\newcolumntype{L}[1]{>{\raggedright\arraybackslash}m{#1}}

\title{NOTSOFAR-1 Challenge: \\ New Datasets, Baseline, and Tasks for Distant Meeting Transcription}

\name{Alon Vinnikov, Amir Ivry, Aviv Hurvitz, Igor Abramovski, Sharon Koubi, Ilya Gurvich, Shai Pe`er, Xiong Xiao, Benjamin Martinez Elizalde, Naoyuki Kanda, Xiaofei Wang, Shalev Shaer, Stav Yagev, Yossi Asher, Sunit Sivasankaran, Yifan Gong, Min Tang, Huaming Wang, Eyal Krupka}

\address{Microsoft}

\email{notsofar\_challenge@microsoft.com}

\begin{document}

\maketitle
 
\begin{abstract}
We introduce the first Natural Office Talkers in Settings of Far-field Audio Recordings (``NOTSOFAR-1'') Challenge\footnote{\url{https://www.chimechallenge.org/current/task2/index}.\label{fn:1}} alongside datasets and baseline system\footnote{A GitHub link will be shared by Feb. 1\textsuperscript{st}.}. The challenge focuses on distant speaker diarization and automatic speech recognition (DASR) in far-field meeting scenarios, with single-channel and known-geometry multi-channel tracks, and serves as a launch platform for two new datasets: First, a benchmarking dataset of 315 meetings, averaging 6 minutes each, capturing a broad spectrum of real-world acoustic conditions and conversational dynamics. It is recorded across 30 conference rooms, featuring 4-8 attendees and a total of 35 unique speakers. Second, a 1000-hour simulated training dataset, synthesized with enhanced authenticity for real-world generalization, incorporating 15,000 real acoustic transfer functions. The tasks focus on single-device DASR, where multi-channel devices always share the same known geometry. This is aligned with common setups in actual conference rooms, and avoids technical complexities associated with multi-device tasks. It also allows for the development of geometry-specific solutions. The NOTSOFAR-1 Challenge aims to advance research in the field of distant conversational speech recognition, providing key resources to unlock the potential of data-driven methods, which we believe are currently constrained by the absence of comprehensive high-quality training and benchmarking datasets.
\end{abstract}

\section{Introduction}

Conversational speech recognition remains a formidable challenge, particularly in meeting scenarios where distant array devices are commonly employed for capturing speech. The distance between speakers and the microphones brings about substantial acoustic complexities including reverberation, variations in speaker distances and volumes, and background noises. These complexities are further amplified by the ever-changing positions of speakers and environmental conditions, altering the acoustic transfer functions (ATFs) \cite{chung1980transfer} between speakers and microphones. Compounding these acoustic challenges are the dynamic aspects of conversations, which include speech overlap, interruptions, rapid changes in speakers, and non-verbal vocalizations \cite{skudrzyk2012foundations}.

With the emergence of Large Language Models (LLMs), conversational speech recognition has been infused with unprecedented practical importance. LLMs fed by speech recognition engines are redefining user-experience with capabilities ranging from meeting summary, note taking and sentiment analysis, to personalized and context-aware responses to user queries. The foundation of these innovative features, particularly in meeting scenarios, lies in distant speaker diarization and automatic speech recognition (DASR) \cite{sellen2023rise}.

The field of conversational speech recognition has seen significant advancements thanks to numerous datasets  such as AMI \cite{kraaij2005ami}, ICSI \cite{dhillon2004meeting}, Sheffield Wargames \cite{fox2013sheffield}, ASpiRE \cite{harper2015automatic}, DIRHA \cite{cristoforetti2014dirha}, VOiCES \cite{richey2018voices}, DiPCo \cite{van2019dipco}, Ego4D \cite{grauman2022ego4d}, AliMeeting \cite{yu2022m2met}, LibriCSS \cite{chen2020continuous}, and challenges such as CHiME \cite{c1,c2,c3,c4,c5,c6,c7}. However, they suffer from various limitations, particularly with respect to advancing DASR. The three most-similar corpora to our work, targeting general meeting scenarios with unsegmented recordings from a single device, are LibriCSS, AMI, and AliMeetings.

LibriCSS makes a significant step forward by including unsegmented semi-real meetings and proposing to measure word error rate (WER) directly rather than signal-based metrics. They also provide matching fully simulated training data for speech separation. However, since their evaluation data has been obtained by playing LibriSpeech utterances via loudspeakers, it does not account for natural conversational dynamics or real-world acoustic complexities, such as time-varying ATFs. AMI makes an impressive contribution collecting 100 hours of real meeting recordings with close-talk annotations. However, its capacity to fully represent real-world complexities is hampered by several limitations. The recordings, confined to just three rooms and no more than four attendees, offer limited acoustic variety. Additionally, the development and evaluation sets consist of only 18 and 16 sessions, respectively. Such a small sample size introduces uncertainty in predicting real-world performance. For instance, a severe diarization error in just one session could significantly impact the overall score. AliMeetings collects a substantial number of real meeting recordings involving a large number of unique speakers. Yet, its focus on Mandarin recordings shifts the challenge more towards language and automatic speech recognition (ASR) aspects, rather than far-field acoustics and conversational dynamics.
Furthermore, their test set consists of only 20 sessions, facing the same small-sample issue as AMI. Notably, neither AMI nor AliMeetings include a matching simulated training set, which hinders the application of data-driven front-ends.

Addressing these gaps and several others, our main contributions are as follows:
\begin{itemize}
    \item \textbf{A meeting dataset for benchmarking and training}: Consisting of roughly 315 distinct meetings and carefully designed for benchmarking purposes. Meetings were recorded across a variety of rooms, while capturing a wide spectrum of real-world scenarios, covering both typical and rare ones, with up to 8 attendees per meeting. Advancing beyond past practices, we emphasize transcription accuracy with a multi-judge annotation process and avoid the use of machine-aid to mitigate related biases. We also include detailed metadata with each meeting for deep-dive analysis, documenting acoustic events and conversational aspects such as the presence of extra overlap. To the best of our knowledge, this dataset is the first to offer this kind of annotation. Furthermore, in our recordings we utilize both single-channel and multi-channel devices, reflecting the range of typical commercial recording equipment.
    \item \textbf{A matching simulated training dataset}: This approximately 1000-hour simulated training set, intended for speech separation and enhancement, improves the authenticity of the simulation process to facilitate the application of data-driven methods. It incorporates 15,000 real ATFs recorded with a geometry matching the multi-channel devices from the real meeting dataset, clean speech is carefully filtered for quality, and the mixing process is designed to exhibit real-world patterns. Additionally, the supervision signals distinguish between the direct and early reflections, and reverb components of an ATF, offering more opportunity for exploration.
    \item \textbf{An open-source baseline system}: To assist participants in getting started, we provide a fully functional baseline system, complete with inference, training, data handling, and evaluation code. The system is written entirely in Python and consists of continuous speech separation (CSS), ASR, and diarization modules \cite{yoshioka2019advances, raj2021integration}.
    \item \textbf{NOTSOFAR-1 Challenge}: We are launching the NOTSOFAR-1 Challenge to pursue fundamental questions in the field. It features single-channel and known-geometry multi-channel tracks, centered on single-device DASR.
\end{itemize}

The rest of this paper is organized as follows: Sections \ref{sec:tracks} and \ref{sec:sc_goals} discuss the challenge tracks, metrics, and scientific goals. Section \ref{sec:data} explores the benchmarking and training datasets and their design considerations in detail. Section \ref{sec:baseline} presents the baseline system. In Section \ref{sec:conclusions}, we conclude.

\section{Tracks and Metrics}\label{sec:tracks}
The challenge features two main tracks centered on the use of a single device: the “single-channel track” and the “known-geometry multi-channel track”, with participants able to submit entries to either one or both. The use of a single device and fixed, known geometry in the multi-channel track are key distinctions from CHiME-8’s DASR challenge\footnote{\url{https://www.chimechallenge.org/current/task1/index}}. 
Participants receive unsegmented recordings from each meeting and are asked to generate time-marked, speaker-attributed transcripts. In line with the notations in \cite{c7}, each meeting is annotated with a set of segments, denoted as $r_i=\left(\Delta,s,\mathbf{v}\right)$, where $\Delta=\left(t_\mathrm{s},t_\mathrm{e}\right)$ are start and end times, $s\in \mathbb{Z}^+$ is the speaker label, and $v\in\Sigma$ is the transcript for the segment, with vocabulary $\Sigma$. A meeting’s reference is fully defined by the $N$ reference segments $R=\{r_1,…,r_N\}$. The participants’ task is to estimate this reference through a generated hypothesis, denoted by $H=\{h_1,…,h_{\hat{N}}\}$. Note that the estimated speaker labels are not required to be identical to the reference labels.

In each track, systems will be ranked according to two metrics: (1) a speaker-attributed metric to evaluate the impact of both speaker diarization errors and word errors, and (2) a speaker-agnostic metric to evaluate word errors. For the first metric we align with CHiME-8's DASR and choose the time-constrained minimum-permutation word-error rate (tcpWER) \cite{MeetEval23}. This is an enhanced version of the concatenated minimum-permutation word-error rate (cpWER) \cite{povey2011kaldi} that incorporates temporal information to prevent matching words that are temporally distant, which better reflects temporal accuracy and speeds up computation compared to cpWER. However, speaker attributed metrics such as tcpWER are highly sensitive to speaker diarization errors: an incorrectly attributed segment will be penalized even if the words within it are accurately recognized. Thus, we include a second speaker-agnostic metric to isolate word error impact, highlighting systems that excel in this aspect.
Additionally, we aim to promote the development of innovative, practical systems rather than performance-squeezing approaches that are more brute-force in nature. We encourage participants to submit their system's runtime and hardware specifications. Systems deemed practical and efficient will be featured on a dedicated leaderboard. 

\section{Scientific Goals}\label{sec:sc_goals}
We aim to address fundamental questions in the field of distant conversational speech recognition:
\begin{itemize}
    \item How much of an advantage does a multi-channel and geometry-specific algorithm have over a single-channel algorithm? 
    \item How much of an advantage, if any, does a geometry-specific algorithm have over a geometry-agnostic algorithm?\footnote{We work closely with CHiME-8's DASR challenge that offers a geometry-agnostic multi-channel track: \url{www.chimechallenge.org/current/task1/index}. Every geometry-agnostic system submitted to their track will automatically yield a result in our geometry-specific track on the NOTSOFAR meeting dataset.}
    \item Can the introduced simulated training dataset lead to data-driven front-end solutions that generalize well to realistic acoustic scenarios? 
    \item How can the various available supervision signals for training be leveraged to improve algorithms? Namely, the separated speech components within the simulated training dataset, along with the close-talk recordings, transcriptions, and speaker labels found in the real meeting dataset.
    \item Can the data verticals analysis enabled by the metadata reveal potential avenues for progress?
\end{itemize}

\section{Training and Benchmarking Datasets}\label{sec:data}

High-quality and high-quantity datasets are critical resources in advancing machine learning research. Particularly in far-field speech processing, we believe the progression of data-driven methods is currently hindered by the absence of comprehensive training and benchmarking datasets. These datasets are crucial for unlocking the full potential of such methods and measuring progress. To address this need, we introduce three key resources: First, a dataset of natural meeting recordings carefully designed for benchmarking purposes. Second, to foster innovation and encourage the adoption of data-driven solutions for speech separation and enhancement, we offer a matched simulated training dataset that bridges several gaps between training and testing conditions. Third, to supplement the simulated training set and to allow training all components of a speech processing pipeline, we allocate a portion of the meeting dataset for training purposes. Both datasets encompass multi-channel and single-channel settings.

\begin{table*}[ht]
\caption{NOTSOFAR Meeting Dataset.}
\centering
\begin{tabularx}{\textwidth}{|>{\raggedright\arraybackslash}m{0.2\textwidth}|Y|Y|Y|}
\hline
\textbf{Feature} & \textbf{Training Set} & \textbf{Development Set} & \textbf{Evaluation Set (blind)} \\
\hline
Number of Meetings & 110 meetings & 35 meetings & 170 meetings \\
\hline
Meeting Duration & 6 minutes & 6 minutes & 6 minutes \\
\hline
Number of SC streams & 5 streams & 5 streams & 6 streams \\
\hline
Number of MC devices & 4 devices & 4 devices & 4 devices \\
\hline
Total Duration of SC Audio Recordings & \thead{55 hours \\ (110 meetings $\times$ \\ 6 minutes $\times$ \\ 5 SC streams)} & \thead{17.5 hours \\ (35 meetings $\times$ \\ 6 minutes $\times$ \\ 5 SC streams)} & \thead{102 hours \\ (170 meetings $\times$ \\ 6 minutes $\times$ \\ 6 SC streams)} \\
\hline
Total Duration of MC Audio Recordings & \thead{44 hours \\ (110 meetings $\times$ \\ 6 minutes $\times$ \\ 4 MC devices)} & \thead{14 hours \\ (35 meetings $\times$ \\ 6 minutes $\times$ \\ 4 MC devices)} & \thead{68 hours \\ (170 meetings $\times$ \\ 6 minutes $\times$ \\ 4 MC devices)} \\
\hline
Total Number of Rooms & 20 rooms & 5 rooms & 13 rooms \\
\hline
Number of Participants & 22 Participants & 11 Participants & 13 Participants \\
\hline
Language & \multicolumn{3}{c|}{English} \\
\hline
\end{tabularx}
\label{table:benchmark}
\end{table*}

\begin{table*}[ht]
\caption{Recording Devices for NOTSOFAR Meeting Dataset.}
\centering
\begin{tabular*}{\textwidth}{|L{0.18\textwidth}|L{0.15\textwidth}|L{0.2\textwidth}|L{0.093\textwidth}|L{0.093\textwidth}|L{0.13\textwidth}|}
\hline
\textbf{Device} & \textbf{Position} & \textbf{Array Geometry} & \textbf{Single-channel stream} & \textbf{Multi-channel stream} & \textbf{Dataset} \\
\hline
Yealink SmartVision 60 & Center of table & Circular 7 Microphones & Yes & Yes & All sets \\
\hline
Yealink Intelligent Speaker & Center of table & Circular 7 Microphones & Yes & Yes & All sets \\
\hline
EPOS Expand Capture 5 & Center of table & Circular 7 Microphones & Yes & Yes & All sets \\
\hline
Sennheiser TeamConnect Intelligent Speaker & Center of table & Circular 7 Microphones & Yes & Yes & All sets \\
\hline
Logitech MeetUp & Front of room & Linear & Yes & No & All sets \\
\hline
Poly Studio & Front of room & Linear & Yes & No & Evaluation set \\
\hline
Jabra PanaCast 50 & Front of room & Linear & Yes & No & Evaluation set (partial) \\
\hline
Shure WH20 + TASCAM DR-40X Close Talk & Head mounted microphone & N/a & Yes & No & All sets (except evaluation set) \\
\hline
\end{tabular*}
\label{table:devices}
\end{table*}

\subsection{NOTSOFAR Recorded Meeting Dataset}
This dataset comprises approximately 315 unique meetings, each lasting on average 6 minutes, featuring authentic, multi-participant English conversations recorded in about 30 different conference rooms at Microsoft offices. To increase acoustic diversity, each meeting was captured with several devices, each positioned differently. This setup, per meeting, involved around 5 single-channel (SC) devices producing a single internally processed stream each, and 4 multi-channel (MC) devices producing 7 raw streams each. In total, this resulted in roughly 170 hours of SC data and 130 hours of MC data across all devices and meetings. We divide the dataset into training (55 SC hours, 44 MC hours), development (17.5 SC hours, 14 MC hours), and evaluation sets (102 SC hours, 68 MC hours). The latter, a subset of which will be used as the fully blind evaluation set in the NOTSOFAR-1 Challenge, is entirely disjoint from the other sets, with no overlap in speakers or rooms. Tables \ref{table:benchmark} and \ref{table:devices} give a compact view of this dataset.
Meticulously designed and collected, the dataset possesses unique characteristics that we believe make it an excellent benchmark for conversational scenarios.

\textbf{Capturing the full spectrum of real-world complexities}: As many practitioners observe, to truly evaluate a system’s accuracy, it is essential to test it not only in typical scenarios but also in rare events that lie at the borders of, or even beyond, its standard operational envelope. In line with this, the recordings encompass a wide array of acoustic situations, including speakers at varying distances and volumes, and different types of transient interference and noises at low and high levels. Importantly, the dataset includes acoustic situations that sometimes substantially and frequently modify the ATF between the active speaker and the device. These range from in-seat movement, walking, standing or sitting, to speaking near a whiteboard, and entering or leaving the room. Conversational dynamics include overlapping speech from multiple speakers, interruptions, rapid speaker change, laughter, coughing, and fillers. The meetings span 30 different rooms of various dimensions, layouts, construction materials and acoustic characteristics. Attendees for each meeting typically vary between 4-8 adults, involving 22 unique speakers in the development and training sets, and a distinct group of 13 speakers in the evaluation set, while ensuring balanced gender representation.

\textbf{Metadata for deep-dive analysis}: The errors of speech recognition systems in conversational settings are notoriously hard to investigate. The multitude of factors impacting the speech pipeline are highly context-dependent and are often undetectable without specialized tooling or manual tagging by human-in-the-loop. To facilitate scalable error analysis and uncovering a system’s blind spots, each meeting is accompanied by metadata, including activities and acoustic events. Some meetings are purposefully dominated by rare events which are documented in the metadata. For example, meetings tagged as extra noisy, or having extra overlapping speech are useful for testing algorithms under those conditions in an isolated manner. Meetings where speakers are in motion or are speaking near sound-reflective whiteboards while turning their heads are useful to test the robustness of an algorithm to changes in the ATF. In the NOTSOFAR-1 Challenge we intend to share the performance evaluations of submissions both averaged across all meetings and within each data vertical.

\textbf{Prioritizing meetings over hours}: Speech datasets often highlight the duration of their recordings in hours, yet the utility of a benchmark is determined by the distribution it spans and the sample size it provides. Recognizing that prolonged recordings from the same meeting offer diminishing returns in terms of yielding new, independent information, our approach focuses on maximizing the diversity of the dataset by including a larger number of distinct meetings. Our dataset consists of approximately 315 relatively short meetings, each lasting roughly six minutes. Since each meeting offers a unique mix of acoustics, speaker identities, and conversational dynamics, we obtain a highly diverse sample. This allows us to assume each meeting is approximately an independent and identically distributed (i.i.d.) element, facilitating the computation of meaningful confidence intervals, a vital aspect of algorithm comparison \cite{payton2003overlapping}. Our baseline code features a simple method for calculating confidence intervals, both in absolute terms and relative to the baseline algorithm. Notably, the narrowness of these confidence intervals is determined by the number of meetings rather than the total recorded hours, and benefits from the large volume of meetings in our dataset.

\textbf{Bias-free transcription process}: Accurate transcription is crucial for training and evaluating speech recognition systems.
One concern in this area is the potential bias toward machine-generated transcriptions. This bias can occur when the annotators are influenced by the output of automated systems, leading to transcriptions that are inadvertently biased toward a particular system’s errors. Consider a machine-aided process where a machine first transcribes the recordings, and a human conducts a second pass, deciding to accept some words and correct others. Due to the inherent ambiguities of speech, even in close-talk recordings, and especially in situations with heated discussions, overlapping speech, and rapid speaker turns, some erroneous machine suggestions may seem reasonable and hence may be accepted by the human annotator. While this process is relatively efficient, it perpetuates errors of the specific transcription algorithm used. The implications of this bias are profound: competing algorithms will be penalized for deviating from the transcription algorithm, potentially concealing the progress of new algorithms. This is particularly problematic when improvements are incremental, as is often the case in research. Since even subtler forms of machine aid can introduce bias, our transcription process for the development and evaluation sets is based on close-talk recordings and conducted entirely without human access to machine-generated transcriptions \cite{levit2017don}.

\textbf{Multi-judge annotation for accurate segmentation and transcription}: To ensure high-quality ground-truth transcription and segmentation, attendees were equipped with close-talk microphones. Spectral subtraction between close-talk channels was used as pre-processing for an automatic segmentation algorithm, followed by human annotators who reviewed and fine-tuned any segmentation errors. Throat microphones were used for quality control and tuning of the algorithms. Professional transcribers carried out the transcription process. Most meetings were annotated independently by two professionals, with a third judge resolving non-consensus cases, except for about 50 meetings in the training set that involved only two annotators. Close-talk recordings of the development and training sets are also released as part of the challenge to encourage exploration of creative methods, such as using pseudo labels. However, these close-talk recordings may still contain a small amount of speech leakage. Hence, to prevent any impact on ASR performance, we recommend masking the close-talk data of each participant by their segmented speech.

\textbf{Distinguishing single-channel from single-microphone}: When evaluating speech recognition systems that process a one-dimensional stream of audio, it is common to use audio recorded from one of the microphones in an array. However, in practice, speech recognition systems often deal with audio from existing conference room devices. These devices are typically equipped with microphone arrays and employ proprietary on-device algorithms to process the audio, yielding what we refer to in this paper as a “single-channel” output. This output is usually the result of processing all the microphones in the array and using a pipeline that includes echo cancellation, de-reverberation, beamforming, and noise suppression. This significantly differs from the single-microphone outcome. To evaluate performance in the single-channel setting, our recordings include a variety of commercially available single-channel devices. Although single-microphone setups are of great scientific interest, particularly when compared to multiple microphones, the NOTSOFAR-1 Challenge's single-channel track focuses on the practical scenario of processing audio from single-channel devices to bridge the gap to real-world applications.

\textbf{A variety of recording devices}: Each meeting was recorded by multiple commercially available conference rooms devices, as listed in Table 2. Each device produced a single-channel stream and most recorded an additional, multi-channel stream. The multi-channel streams are recorded in raw format directly from the microphone arrays without processing and their geometry is of one central and six surrounding microphones, matching the simulated training set described in the next section. The purely single-channel devices have linear microphone-arrays with different numbers of microphones. In the NOTSOFAR-1 Challenge, although multiple devices were used for recording, processing during inference is restricted to just one of these devices. The head-mounted close-talk microphones featured a unidirectional cardioid pickup pattern that provides some isolation of each speaker from their nearby speakers, allowing for more faithful transcription even in the presence of loud overlapped speech. In every conference room, we played chirps (frequency sweeps) with a loudspeaker and recorded them with each of the multi-channel devices. These recordings are supplied as well and may be used to compute reverberation times, e.g., RT60 and T30 \cite{matt2019estimating}.

\textbf{Oriented toward far-field}: The primary challenge in far-field speech transcription lies in creating an algorithm that is robust to the acoustic complexities introduced by the distance between the microphone and the speaker, as well as the surrounding environmental conditions. Some aspects of our dataset are designed for far-field evaluation. Namely, we deliberately steer away from challenges related to different languages, accents, and domain-specific jargon. The recordings are exclusively in English, spoken by individuals who are either native speakers or possess near-native proficiency. The language employed is predominantly semi-professional. While the meetings encompass a wide range of topics, care was taken to avoid specialized jargon or obscure terminology, thereby focusing more on the acoustic challenges rather than linguistic diversity.

\textbf{A multitude of topics and interaction styles}: Most meetings featured semi-professional topics, in which participants role-played as professionals discussing a work-related issue. For example, a cruise ship company planning an event, administrators planning a city park, or users complaining about IT problems. Some meetings featured non-work-related topics, such as favorite TV shows, debating whether to raise kids as vegetarians, or friends sharing recipes. We expect this role-playing to create a range of conversational dynamics akin to those found in actual meetings in terms of speech overlap, turn-taking dynamics, speech disfluencies, interruptions and so on.

\subsection{NOTSOFAR Simulated Dataset}
The simulated training set consists of about 1000 hours simulated with the same microphone-array geometry of one central and six surrounding microphones, matching the multi-channel devices in the NOTSOFAR meeting dataset. It provides separated speech and noise components as supervision signals for training data-driven speech separation and enhancement methods. It is designed to foster generalization to realistic settings by identifying and closing major train-test gaps prevalent in simulated datasets in prior works.

\textbf{Real-room acoustic transfer functions}: Simulated datasets for speech separation typically rely on the Image Method \cite{allen1979image} to generate simulated ATFs, which are then convolved with clean speech to further simulate in-room utterances. Despite advancements in simulation techniques, simulated ATFs do not take into account many elements present in the real-world, such as: the room’s layout and the unique reflective properties of its surfaces, the dependency of the reflection coefficients on frequency and angle of incidence, the reflections from adjacent speakers, the directivity of the speech source, and the acoustic effects of the specific microphone array casing. The limitations of the Image Method in this context have been explored in several studies, e.g., \cite{savioja2015overview, brinkmann2019round, de2015modeling}. To address these shortcomings, our dataset features ATFs recorded in actual conference rooms, in an acoustic setup arranged to closely replicate authentic meeting environments. We collected a total of 15,000 real ATFs, measured in various positions and rooms by multiple devices sharing the same geometry. The ATFs were reconstructed from chirps emitted by a mouth simulator. 

\textbf{MOS-filtered clean speech}: For clean speech we use the Librivox corpus \cite{kearns2014librivox}, which contains recordings of varying quality. Many of these recordings are of excellent speech quality, but others exhibit poor quality in the form of speech distortion, background noise and reverberation. This quality variance compromises simulation authenticity, since the ATF convolution operation assumes true clean speech. Hence, we follow the work of \cite{reddy2020interspeech} and choose the upper quartile with respect to the subjective mean-opinion score (MOS) as our clean speech corpus, amounting to 500 hours of speech.

\textbf{Augmented clean speech for realistic overlap patterns}: We observe that the Librivox clean speech corpus, comprising recordings of individuals reading audiobooks, typically consists of rather continuous segments of speech lasting several seconds, without significant pauses. This makes it harder to achieve a mixture reflecting realistic overlap patterns which include short interruptions and rapid speaker turns. We therefore apply a signal processing algorithm that detects drops in speech power to identify short pauses in speech and subsequently insert random-length silence breaks at these points. By augmenting clean speech in this manner, we facilitate a more realistic simulation of conversational dynamics during the mixing stage.

\textbf{Mixing up to three speakers}: During the mixing process we first convolve the augmented clean speech utterances with different ATFs selected from the same room. The resulting individual utterances are roughly 40 seconds long. We then apply a small random shift before summing utterances from 3 distinct speakers. This strategy results in mixtures that average 50 seconds in length and possess two attractive characteristics. First, the mixtures display real-world patterns such as complex overlaps, and rapid speaker turns featuring up to three distinct speakers. Qualitatively we observe that randomly sampled segments of 2-10 seconds, as commonly utilized in speech separation training, effectively capture patterns that are typical in real meetings. Second, these segments tend to be more challenging, serving as a form of hard sample mining, which is beneficial for training data-driven methods. Finally, transient and stationary noises recorded in real rooms with the matching geometry were added to the mixture.

\textbf{Separating direct and early reflections from late reverberations}: Previous studies have highlighted that speech signals captured by a distant microphone are generally smeared by reverberation, which severely degrades automatic ASR performance \cite{lebart2001new}. Motivated by this as well as internal experimentation findings, we have designed the supervision signals in this training set to facilitate not only speech separation training but also dereverberation. We assume a room impulse response (RIR) consists of three parts: a direct-path response, early reflections, and late reverberations. Since late reverberations are known to be a major cause of ASR performance degradation, we divide the recorded ATFs (now viewed as RIRs) into two components: direct and early reflections, and late reverberations. This division is accomplished by applying a smoothed-out cutoff at the 50-millisecond mark. 
The mixed signal at time \( t \), with \( i \) indexing speakers, is then represented as follows:
{
\begin{align}
   \mathrm{mixture}(t) &= \sum_{i} \left[ \mathrm{spk}_{i}^{\textnormal{direct+early}}(t) + \mathrm{spk}_{i}^{\textnormal{reverb}}(t) \right] + \mathrm{noise}(t), \nonumber
\end{align}
}The individual ground truth (GT) components are provided as supervision signals in our dataset. The baseline model in the NOTSOFAR-1 Challenge employs the direct and early reflections component as target labels for its speech separation module. However, since the sensitivity to late reverberations depends on the specific ASR system, participants are encouraged to construct their own target labels, possibly incorporating the full RIR.

\textbf{Limitations}: A notable limitation of the simulated training set is the static nature of the ATFs, which, despite being recorded in real rooms, remain fixed throughout each simulated utterance. This fails to mirror common real-world situations where the ATF varies in time as the speaker or environment moves. To evaluate an algorithm's robustness against time-varying ATFs, one can leverage the metadata provided in the NOTSOFAR meeting  dataset. 
Furthermore, we hope that the training portion of these real recordings, which includes instances of time-varying ATFs, will contribute significantly to improving an algorithm’s robustness to such variations.

\section{Baseline System}\label{sec:baseline}
The baseline recipe of NOTSOFAR-1, inspired by \cite{yoshioka2019advances, raj2021integration}, consists of three steps: continuous speech separation (CSS), ASR, and speaker diarization. In this section, we will briefly describe each step. 

\textbf{CSS}: The objective of CSS is to receive an input audio stream, consisting of either 1 or 7 channels, and convert it into a set of overlap-free streams. In our case, we generate $N=3$ distinct speech streams, which can support up to 3 speakers overlapping at the same time. An effective CSS essentially monitors the input audio stream, and when overlapping utterances are found, it distributes them to the different output channels. We follow the conventional CSS framework, first training a speech separation network with permutation invariant training (PIT) loss, which takes fixed-length audio segments as input and outputs $N$ speech masks and one noise mask. The masks are in STFT (short-time Fourier transform) domain \cite{durak2003short}. For the network architecture we select the conformer model \cite{chen2020continuous, gulati2020conformer}. 

In the inference phase, the network is applied in a block-wise streaming fashion to overlapping fixed-length segments. Since the order of the $N$ speech mask outputs may not be consistent across segments, we align every pair of adjacent segments. To estimate the best order, we consider all permutations and select the one with the lowest mean squared error (MSE) between the masked magnitude spectrograms, calculated over the frames shared by the two segments. After stitching the $N$ speech masks and one noise mask over time, we proceed to generate each of the N output streams. For the single-channel variant, this consists of multiplying the input mixture by the target speech mask. For the multi-channel variant, we rely on mask-based minimum variance distortionless response (MVDR) beamforming \cite{habets2010mvdr}. As part of this scheme, the spatial covariance matrices (SCMs) of the target and interference signals are computed, where the interference signal is defined as the sum of all non-target speakers and the background noise.

\textbf{ASR}: For automatic speech recognition (ASR), we employ Whisper “large-v2” \cite{radford2022robust}, which supports word-level timestamps. ASR is applied independently to each audio stream produced by CSS.

\textbf{Speaker Diarization}: The task of the diarization module is to assign a speaker label to every word transcribed by the ASR module. Speaker labels are unique identifiers such as `spk0', `spk1', etc., and they are not required to be identical to the reference speakers. We first apply an offline diarization method to diarize each of the audio streams produced by CSS. Then, we assign each word to its speaker label based on its word boundary information and the diarization output from the source stream.

For offline diarization, we adopted the diarization recipe of the NeMo toolkit \cite{kuchaiev2019nemo}. There are two configurations supported, including “nmesc” which performs offline clustering using the normalized maximum eigengap-based spectral clustering (NME-SC) algorithm \cite{park2019auto}, and “nmesc-msdd” which performs NME-SC followed by the multi-scale diarization decoder (MSDD) \cite{park2022multi}. 

To assign a speaker label to an ASR word, we look up the active speakers within its time boundaries in the diarization output from the corresponding audio stream. In most cases, there is only one active speaker within the word’s boundaries, and it is assigned to the word. If there is no active speaker within the word’s boundaries (i.e. diarization didn’t detect speech), the speaker label of the nearest word in time is assigned. If there are multiple active speakers within the word’s boundaries (i.e. diarization detected overlapped speech), the speaker who is active for the longest duration is assigned. 

Note that the sole distinction between the single-channel and multi-channel variants of our system is in the CSS module which processes either 1 channel or 7 channels as its input, applying mask multiplication or MVDR respectively.

\section {Conclusions}\label{sec:conclusions}
The NOTSOFAR-1 Challenge marks a significant stride in the field of DASR. It provides a 315-meeting benchmarking dataset that captures the nuances and complexities of real-world meetings. The benchmark includes a large volume of distinct meetings, enabling reliable performance analysis of algorithms. Annotations were collected by human listeners, providing ground-truth annotations with low WER. The challenge also includes a 1000-hour simulated training data set, crafted to leverage known microphone array geometry, and mitigate the mismatch between training and testing conditions. These contributions collectively aim to stimulate further research and innovation, and to push the boundaries of what is scientifically possible in the field of DASR.

\section{Acknowledgements}
We express our profound gratitude to Hadas Eilon Carmi, Ross Cutler, Prem Premkumar, Solomiya Branets, and the IC3-AI team at Microsoft for their invaluable assistance.

\bibliographystyle{IEEEtran}
\bibliography{mybib}

\end{document}